\documentclass[aps,prl,twocolumn,groupedaddress]{revtex4}
\usepackage{graphicx}
\usepackage{bm}

\begin{document}

\title{Double layer two-dimensional electron systems: Probing the transition 
from weak to strong coupling with Coulomb drag
}

\author{M. Kellogg$^1$, J.~P. Eisenstein$^1$, L.~N. Pfeiffer$^2$, and K. W. 
West$^2$}

\affiliation{$^1$California Institute of Technology, Pasadena CA 91125 
\\
	 $^2$Bell Laboratories, Lucent Technologies, Murray Hill, NJ 
07974\\}


\begin{abstract} 
Frictional drag measurements revealing anomalously large dissipation at the 
transition between the weakly- and strongly-coupled regimes of a bilayer two-dimensional electron system at total Landau level filling factor $\nu_T =1$ are 
reported.  This result suggests the existence of fluctuations, either static or 
dynamic, near the phase boundary separating the quantized Hall state at small 
layer separations from the compressible state at larger separations. Interestingly, the anomalies in drag seem to persist to larger layer separations 
than does interlayer phase coherence as detected in tunneling.   

\end{abstract}


\maketitle

At high magnetic fields two-dimensional electron systems (2DES) exhibit a 
variety of collective states.  For example, if the perpendicular 
magnetic field is adjusted so that the density of electrons $N_s$ equals one-half the degeneracy $eB/h$ of the lowest spin resolved Landau level (i.e. at 
filling factor $\nu =N_sh/eB =1/2$), the resulting strongly correlated electron 
system can be successfully modeled as a metallic liquid of composite 
fermions\cite{halperin}.  The system crudely resembles a conventional 2DES in 
zero magnetic field. No quantized Hall effect is seen. Remarkably, the 
system possesses a well-defined Fermi surface and quasiparticles which move in 
semiclassical cyclotron orbits whose diameters diverge as $\nu \rightarrow 1/2$.  

Now consider a system consisting of two parallel 2DESs, each at $\nu=1/2$, 
separated by a barrier layer.  Obviously, if the separation $d$ between the two 
2DESs is large, they are uncoupled and the net system behaves much the same as a 
single layer.  In contrast, if $d$ is very small (less than the average 
separation between electrons in each layer) the ground state of the bilayer 
system is qualitatively different.  In this strongly-coupled limit interlayer 
Coulomb interactions engender a novel broken symmetry, spontaneous interlayer 
phase coherence, in which all electrons are coherently spread between both 
layers, even in the hypothetical limit of zero interlayer 
tunneling\cite{girvinmacd}.  This bizarre state, which is best characterized by 
the {\it total} filling factor $\nu_T=1$, may be viewed in a number of 
equivalent ways, including as an itinerant pseudospin ferromagnet or a Bose 
condensate of interlayer excitons.  Experimentally, the system has been found to 
display numerous striking properties, including the quantized Hall effect (at 
$R_{xy}=h/e^2$)\cite{jpe1}, a pseudospin textural phase 
transition\cite{murphy,yang}, Josephson-like interlayer 
tunneling\cite{spielman}, and, most relevant here, quantized Hall 
drag\cite{kellogg}.  Additional properties, including counterflow superfluidity, 
are anticipated\cite{supermac}.

The nature of the transition between the strongly-coupled ferromagnetic or 
excitonic phase at small $d$ and the weakly-coupled composite fermion liquids at 
large $d$ is very poorly understood.  As $d$ increases from zero (at zero 
temperature) the ferromagnetic phase suffers increasingly severe quantum 
fluctuations which eventually destroy the (algebraic) pseudomagnetic order at a 
critical layer separation.  Although there exists numerical evidence suggesting 
that this quantum phase transition may be weakly first order\cite{schliemann}, 
current experiments suggest a continuous transition.   

Beyond the nature of the demise of the ferromagnetic phase as $d$ increases, 
remains the question of the ground state of the system above the critical 
separation.  Before reverting to independent composite fermion liquids at very 
large $d$, there is the possibility of additional interlayer correlated phases 
at intermediate separations.  Candidate states include bilayer Wigner 
crystals\cite{shayegan}, paired composite fermion liquids\cite{bonesteel}, and 
various other exotic phases which may share some but not all of the properties 
of the ferromagnetic phase (e.g. interlayer phase coherence but no quantized 
Hall effect, etc.)\cite{demler,kim}.  In this paper we report interlayer 
friction measurements (``Coulomb drag'') which shed light on the transition 
between the strongly- and weakly-coupled regimes at $\nu_T=1$.  In particular, 
we observe a large enhancement of the longitudinal component of the drag near 
the transition.  Although rapidly attenuated, this enhancement persists to 
surprisingly large layer separations.

Drag measurements are performed by driving a current $I$, typically 1nA at 5Hz, 
through one of the two layers of a double layer 2DES while monitoring the 
voltage $V_D$ which appears in the other, electrically isolated, layer.  At zero 
magnetic field the drag resistance $R_D=V_D/I$ provides a unique measure of the 
interlayer momentum relaxation rate.  For closely-spaced layers and low 
temperatures this rate is dominated by simple Coulomb scattering and hence the 
moniker Coulomb drag.  

The samples used in the present experiments consist of two 18nm GaAs quantum 
wells separated by a 10nm barrier layer of $\rm Al_{0.9}Ga_{0.1}As$.  Each 
quantum well contains a 2DES which, in the sample's as-grown state, has a 
density of about $N_s \approx \rm 5.3\times 10^{10} cm^{-2}$ and a low-temperature mobility of $\mu \approx \rm 10^6 cm^2/Vs$.    Data from three 
samples, A, B, and C, are reported here.  Samples A and B consist of square 
mesas, $250 \mu \rm m$ on a side, with four arms extending outward to remote 
ohmic contacts.  Sample C is bar-shaped, $40\times 400 \mu \rm m$, and has five 
arms and ohmic contacts.  These contacts are connected to one or the other 2DES 
using a selective depletion scheme\cite{jpe}.  The densities of the individual 
2DESs in the central mesa region are controlled with electrostatic gates 
deposited on the top and backside of the samples. A detailed study of the 
Coulomb drag in these samples at zero magnetic field has been reported 
elsewhere\cite{kellogg2}.

\begin{figure}
\includegraphics[width=3.25 in]{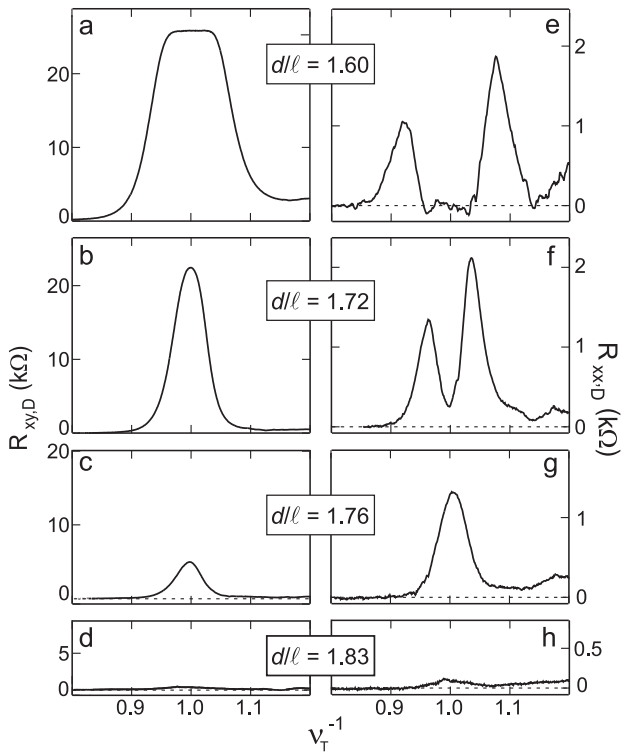}
\caption{Coulomb drag resistances in sample A at $T=30 \rm mK$, in the vicinity of $\nu_T=1$. Left column: Hall drag $R_{xy,D}$; right column: longitudinal drag 
$R_{xx,D}$. Each row corresponds to a different 2D density in the bilayer 
sample, and thus a different effective layer separation $d/\ell$ at $\nu_T=1$.
}
\end{figure}
Figure 1 shows representative Coulomb drag data from sample A in the vicinity of 
$\nu_T=1$ at $T=30 \rm mK$.  Each row corresponds to a different effective layer 
separation $d/\ell$, with $d=28{\rm nm}$ the center-to-center quantum well 
separation and $\ell=(\hbar/eB)^{1/2}$ the magnetic length at $\nu_T=1$. This 
key ratio governs the relative importance of inter- and intra-layer Coulomb 
interactions in the bilayer system, and can be varied {\it in situ} by 
symmetrically changing the electron densities in the individual quantum wells.  
The data in Fig. 1 is therefore plotted versus the inverse total filling factor 
${\nu_T}^{-1}$, instead of magnetic field, to aid in comparing the different 
densities.  The right-hand panels display the longitudinal drag resistance 
$R_{xx,D}$ (i.e. drag voltage parallel to the current flow in the drive layer) 
while the left-hand panels present the transverse, or Hall, drag resistance 
$R_{xy,D}$.    

The top two panels ($a$ and $e$) of Fig. 1 show Hall and longitudinal drag data 
at $d/\ell=1.60$.  At this effective layer separation the system is well within 
the strongly-coupled bilayer quantum Hall effect phase and, as reported 
previously\cite{kellogg}, at $\nu_T=1$ the Hall drag exhibits a plateau 
accurately quantized at $R_{xy,D}=h/e^2$ while the longitudinal drag $R_{xx,D}$ 
is essentially zero.  While it is not surprising that $R_{xx,D}$ is zero in the 
quantized Hall state (the gap to charged excitations suppressing dissipation at 
low temperatures), the quantization of Hall drag is a dramatic consequence of 
the physics of the strong coupling regime and supports, albeit indirectly, the 
existence of counterflow superfluidity\cite{supermac}. 

On moving away from $\nu_T=1$, the Hall drag rapidly diminishes while the 
longitudinal drag displays two strong maxima.  The sign of the Hall drag voltage 
is the same as that of the conventional Hall voltage in the current-carrying 
layer.  In contrast, the sign of the longitudinal drag voltage is the {\it 
opposite} of the conventional longitudinal voltage drop in the current-carrying 
layer. Indeed, for all the data reported here the sign of $R_{xx,D}$ is the same 
as that encountered at zero magnetic field where the drag voltage reflects the 
electric field needed to counteract the frictional force due to the current 
flow in the drive layer.

The remaining panels of Fig. 1 show how Coulomb drag changes as $d/\ell$ 
increases and the strongly-coupled bilayer quantized Hall phase collapses.
Panels $b$, $c$, and $d$ show that the Hall drag plateau becomes a rapidly 
subsiding local maximum as the $d/\ell$ increases.  More interestingly, panels 
$f$, $g$, and $h$ reveal that the longitudinal drag first evolves from a broad 
zero into a local minimum between tall peaks.  By $d/\ell=1.76$ these peaks have 
merged to form a single peak. Further increases of $d/\ell$ steadily reduce the 
magnitude of this peak in $R_{xx,D}$.

Figure 2 displays both drag resistances precisely at $\nu_T=1$ as functions of 
$d/\ell$. These data were obtained from sample B at $T=50 \rm mK$.  As the 
figure indicates, the Hall drag resistance $R_{xy,D}$ undergoes a rapid yet 
smooth transition from very small values above $d/\ell \approx 1.8$ to the very 
large value of $h/e^2=25.8 \rm k \Omega$ for $d/\ell$ below about 1.65.  At the 
same time, the longitudinal drag $R_{xx,D}$ exhibits a strong and rather 
symmetric peak in the transition region.  At $T=50 \rm mK$ this peak is centered 
at $d/\ell \approx 1.73$ and has a half-width of about $\Delta(d/\ell)=0.035$.  
The height of the peak, about $1.8 \rm k \Omega$, represents an impressively 
large drag resistance.  This value is in fact roughly comparable to the 
conventional longitudinal resistance $R_{xx}$ of the bilayer system under the 
same conditions (the deep minimum in $R_{xx}$ characteristic of the $\nu_T=1$ 
QHE developing only at lower values of $d/\ell$). Both the $d/\ell$ location and 
the height of this peak in $R_{xx,D}$ vary slightly from one sample to the next, 
but its existence and qualitative behavior is quite robust.

\begin{figure}
\centering
\includegraphics[width=3.25in]{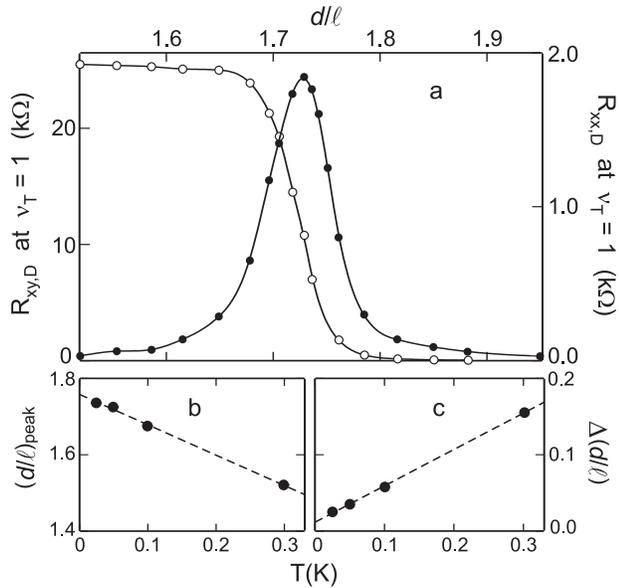}
\caption{a) Hall (open dots) and longitudinal (closed dots) drag at $\nu_T=1$ 
and $T=50{\rm mK}$ $vs.$ $d/\ell$ in sample B. b) and c) Temperature dependence 
of location and half-width of peak in $R_{xx,D}$ at $\nu_T=1$. Lines are guides to the eye.}
\end{figure}

The width and location of the peak in $R_{xx,D}$ at $\nu_T=1$ depend upon 
temperature.  Figure 2b reveals that the peak moves to lower $d/\ell$ as $T$ 
increases.  This dependence, which is roughly linear in $T$, extrapolates to 
about $d/\ell =1.76$ as $T \rightarrow 0$.  At the same time, Fig. 2c shows that 
the peak half-width, $\Delta(d/\ell)$, increases substantially as the 
temperature is increased to 300mK. Interestingly, over the range $25 {\rm mK} 
<T< 300 {\rm mK}$ the height of the peak in $R_{xx,D}$ at $\nu_T=1$ varies by 
only about 15\%.

Figure 3 displays the temperature dependence of $R_{xx,D}$ at $\nu_T=1$ at three 
different values of $d/\ell$.  For $d/\ell=1.58$, which is well within the 
bilayer QHE, $R_{xx,D}$ rises as the temperature is reduced 
from $T=0.5 \rm K$ to about 0.2K but then drops precipitously as $T$ is further 
reduced.  In this low temperature regime $R_{xx,D}$ is thermally 
activated\cite{kellogg}, i.e. $R_{xx,D} \sim e^{-E_A/T}$, with $E_A \approx 0.4K$.  This dependence is expected within the gapped QHE phase and, 
indeed, the conventional resistivity $R_{xx}$ shows the same activation 
energy\cite{kellogg}.  At $d/\ell =1.93$, which is in the non-QHE compressible 
phase, $R_{xx,D}$ is much smaller in magnitude and falls monotonically as $T$ is 
reduced.  This temperature dependence is reminiscent of that seen at $B=0$ in the present samples\cite{kellogg2} and at $\nu_T=1$ in much 
more widely-spaced ($d/\ell \approx 3.9$) double layer 2D electron 
systems\cite{lilly}. Roughly speaking, this behavior reflects the characteristic 
reduction of the phase space for inelastic Coulomb scattering events as the 
temperature falls.  A quantitative model for Coulomb drag at $\nu_T=1$ and 
large $d/\ell$ has been developed by Ussishkin and Stern\cite{ady}.

\begin{figure}
\centering
\includegraphics[width=3.25in]{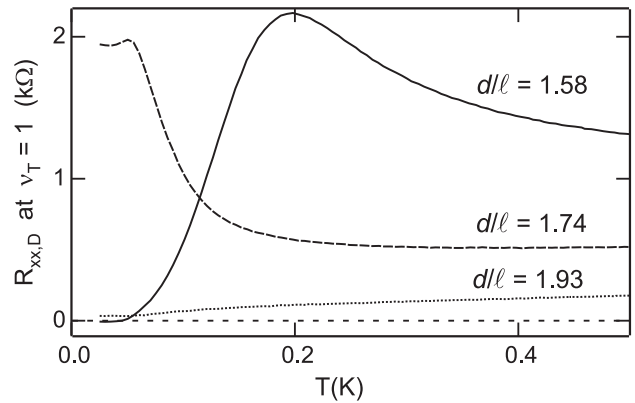}
\caption{Temperature dependence of longitudinal drag resistance at $\nu_T=1$ in 
sample B at three different $d/\ell$ values.}  
\end{figure}

At $d/\ell=1.74$, i.e. in the middle of the transition region, the temperature 
dependence of $R_{xx,D}$ at $\nu_T=1$ is markedly different than at both larger 
and smaller $d/\ell$.  After remaining roughly constant on cooling from $T=0.5 
\rm K$ to about 0.2K, the drag rises rapidly as the temperature falls further. 
Below about $T=50 \rm mK$ $R_{xx,D}$ apparently saturates. 

The data described above show that Coulomb drag is a sensitive indicator of the 
transition between the weakly and strongly-coupled regimes of a bilayer 2DES 
at $\nu_T=1$. This is especially clear from the behavior of the 
Hall drag resistance $R_{xy,D}$ shown in Fig. 2. Although non-zero Hall drag 
can in principle result from density (or energy) dependent 
scattering rates in a 2D system\cite{hu,vonOppen}, the development of a large, 
and ultimately quantized, $R_{xy,D}$ is generally believed to require non-perturbative interlayer correlations.  Indeed, the qualitative behavior of 
$R_{xy,D}$ shown in Fig. 2 was anticipated\cite{renn,moon,duan,kun,kunmac}.  

In contrast, the strong peak in the $\nu_T=1$ longitudinal drag resistance 
$R_{xx,D}$ which develops in the middle of the Hall drag transition comes as a 
surprise.  It seems reasonable to interpret the $d/\ell$ value at the center of 
this peak as the critical one separating the strongly-coupled $\nu_T=1$ QHE 
phase from the weakly-coupled non-QHE phase.  Within the ferromagnetism picture 
of the QHE phase, at zero temperature this critical point marks the destruction 
of the ordered state by quantum fluctuations of the pseudospin moment. The 
shifting of the peak to lower $d/\ell$ as the temperature rises (cf. Fig. 2b) is 
consistent with thermal fluctuations further destablizing the ordered state. The 
non-zero width of the peak in $R_{xx,D}$ $vs.$ $d/\ell$ suggests an 
inhomogeneous situation in which the 2D electron system fluctuates between the 
QHE and non-QHE phases. Stern and Halperin (SH)\cite{adybert} have suggested 
that these fluctuations are static and result from mesoscopic spatial 
inhomogeneities of the 2D electron density.  On the other hand, dynamic critical 
fluctuations in an otherwise homogeneous system could also be involved. 

In the SH picture, as $d/\ell$ is reduced toward the critical value puddles of 
the strongly-coupled QHE phase appear within a background of weakly-coupled non-QHE fluid.  As $d/\ell$ is reduced further, these puddles eventually percolate. 
Via an analysis which assumes the puddles are (as expected) counterflow 
superfluids while the background fluid is a conventional double 
layer 2D conductor with a large Hall resistance but little Coulomb drag, SH 
conclude that the macroscopically averaged longitudinal drag resistivity 
$\rho_{xx,D}$ of the composite system can become very large just before the 
puddles percolate.  In this situation, SH predict that $\rho_{xx,D}$ grows as 
the temperature falls, eventually saturating near $h/2e^2 \approx 13 {\rm 
k}\Omega$.  Figure 3 demonstrates that such a qualitative temperature dependence 
is observed.  Although $R_{xx,D}$ near the midpoint of the transition region 
reaches only $\approx 1.9 {\rm k}\Omega$ as $T \rightarrow 0$, classical 
geometric effects\cite{vdp} associated with our square sample geometry suggest 
that the experimental longitudinal drag {\it resistivity} may be as much as a 
factor of $\pi/ln2$ larger than $R_{xx,D}$, or about $8.6 {\rm k}\Omega$. Further experiments may reveal whether, as SH suggest, the remaining discrepancy is due to the finite conventional (i.e. parallel transport) resistivity of the sample at $\nu_T=1$.

\begin{figure}
\includegraphics[width=3.25in]{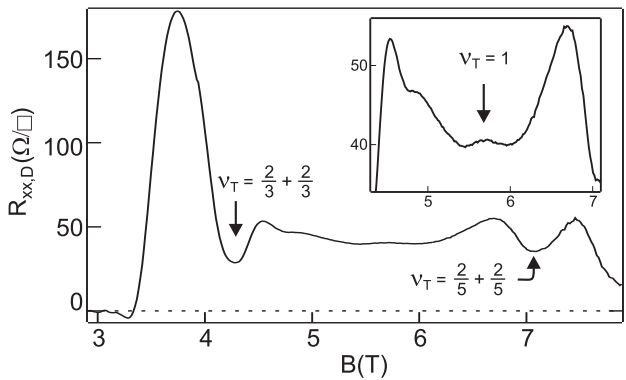}
\caption{Longitudinal drag in sample C at $T=300{\rm mK}$. Expanded view in the 
inset reveals a small enhancement near $\nu_T=1$ where $d/\ell=2.6$.}
\end{figure}

We turn finally to the behavior of Coulomb drag at larger layer separations.
Prior experiments\cite{lilly}, at $d/\ell\approx 3.9$, showed no evidence of any 
anomaly in $R_{xx,D}$ at {\it total} Landau level filling $\nu_T=1$.  As Fig. 2 
shows, the enhancement of $R_{xx,D}$ at $\nu_T=1$ in the present samples 
subsides rapidly as $d/\ell$ increases.  Surprisingly, however, it remains 
observable out to $d/\ell\approx 2.6$.   At these large $d/\ell$ the enhancement 
appears as a small bump on top of a background arising from drag scattering 
processes between weakly-coupled 2D layers - see Fig. 4. The bump at $\nu_T=1$ 
is a genuine bilayer effect: It remains present even when small antisymmetric 
density imbalances are imposed on the double layer system.  By contrast, the 
other features (e.g. the minima at $\nu_T=2/3+2/3$ and 2/5+2/5) seen in Fig. 4 
split in two, thus proving that they are fundamentally single layer effects. 

The existence of enhanced longitudinal drag at $\nu_T=1$ at such large $d/\ell$ 
is not understood.  In principle, there may be local regions in our samples in 
which $d/\ell$ has been reduced by atomic steps in the various heterointerfaces.  
However, the needed reduction is $\sim 30\%$, or about 8nm, and this seems implausibly large\cite{steps}.  We emphasize that no analogous anomaly is seen in the zero bias interlayer tunneling conductance.  The enhanced tunneling, which heralds the onset of interlayer phase coherence\cite{spielman}, is either absent or unobservably small for $d/\ell \ge 1.85$.  This discrepancy raises the 
possibility that the enhanced $\nu_T=1$ drag at larger $d/\ell$ may not require 
interlayer phase coherence. 

We thank Ady Stern, Bertrand Halperin and Steven Girvin for enlightening 
discussions and Ian Spielman for help with sample fabrication. This work was 
supported by the NSF under Grant No. DMR0070890 and the DOE under Grant No. DE-
FG03-99ER45766.


\begin{thebibliography}{99}
\bibitem{steps} Fluctuations in the layer dimensions might, however, contribute to the apparent $T\rightarrow 0$ width of the peak in $R_{xx,D}$ $vs.$ $d/\ell$ seen in Fig. 2c.
\end{thebibliography}
\end{document}